\documentclass[11pt]{article}
\usepackage{hyperref}
\hypersetup{
pdftitle={The Bayesian Who Knew Too Much},
pdfauthor={Yann Ben\'etreau-Dupin},
colorlinks=true,
linkcolor={black},
citecolor={blue},
urlcolor={blue},
}
\usepackage{setspace,commath,amsfonts,natbib,geometry,lastpage,fancyhdr,wrapfig,graphicx}
\usepackage{graphicx,commath,amsfonts,yfonts}
\newcommand{\ud}{\,\mathrm{d}}

\begin{document}

\title{\textbf{The Bayesian Who Knew Too Much}\footnote{The final publication will be available at Springer via \href{http://dx.doi.org/10.1007/s11229-014-0647-3}{http://dx.doi.org/10.1007/s11229-014-0647-3}. The accepted version of this paper is available at \href{http://publish.uwo.ca/~ybenetre/Research_files/BWKTM_Preprint.pdf}{http://publish.uwo.ca/$\sim$ybenetre/Research\_files/BWKTM\_Preprint.pdf}. Section~\ref{sec:10} is an excerpt of \citep{Benetreau-Dupin2015}.}}


\author{\textbf{Yann Ben\'etreau-Dupin}\footnote{Dept of Philosophy \& Rotman Institute of Philosophy, Western University, London, Ontario, Canada. \href{mailto:ybenetre@uwo.ca}{ybenetre@uwo.ca}}}



\date{\small{Preprint, December 2014. Forthcoming in \emph{Synthese}.}}

\maketitle

\begin{abstract}

In several papers, John Norton has argued that Bayesianism cannot handle ignorance adequately due to its inability to distinguish between neutral and disconfirming evidence. He argued that this inability sows confusion in, e.g., anthropic reasoning in cosmology or the Doomsday argument, by allowing one to draw unwarranted conclusions from a \emph{lack of knowledge}. Norton has suggested criteria for a candidate for representation of neutral support. Imprecise credences (families of credal probability functions) constitute a Bayesian-friendly framework that allows us to avoid inadequate neutral priors and better handle ignorance. The imprecise model generally agrees with Norton's representation of ignorance but requires that his criterion of self-duality be reformulated or abandoned.
\vspace{1em}
\end{abstract}

\section{Introduction: Bayesian Reasoning and Ignorance}
\label{intro}

If you ask me what the probability is of rolling a 2 with a throw of a fair, cubic die, I will answer, ``$\frac{1}{6}$.'' Now imagine that we are having a conversation about Shakespeare's play \emph{The Tempest} and you ask me to assign a probability to the claim that the play has 16,633 words. My response would be that I simply do not have a clue. My knowledge of Elizabethan drama would allow me to say that the value falls within some interval, in which all values seem equally plausible. However, I would not be able to assign a probability to the claim that the play has \emph{exactly} 16,633 words. If you insist on a probability, I would respond---echoing David Albert---by asking, ``What part of `I don't have a clue' do you not understand?''

\par The simplest version of Bayesianism is ill-equipped to handle such a case. Reasoning with the principle of indifference, as in the case of the die, motivates a uniform probability distribution over some interval of values; if there are 20,000 equiprobable numbers in the plausible range of number of words for that play, each number will be assigned a probability of $\frac{1}{20,000}$. Thus according to usual Bayesian confirmation theory, learning the exact length of \emph{The Tempest} is equivalent to confirming a claim with a very low probability. In contrast, suppose that we are discussing whether or not the 1956 movie \emph{Forbidden Planet} is a good adaptation of \emph{The Tempest}, and you ask me to assign a probability to the claim that Shakespeare himself authored that movie's screenplay. I would \emph{not} be indifferent about that claim; I would assign to it a very low probability, and I would be very surprised if it were confirmed. Yet orthodox Bayesianism does not allow one to distinguish these two confirmations: it sees them both as the confirmation of a low-probability claim, thereby justifying in both cases a same sense of surprise. One may legitimately be surprised by the confirmation of a low-probability proposition, but not by the confirmation of something about which we are entirely \emph{indifferent}. In the absence of a representation of \emph{neutral} degree of belief, we have no choice but to treat large numbers of alternatives about which we are equally \emph{indifferent} as if they all were \emph{improbable} propositions. Consequently, any hypothesis---however unlikely---that lends more support to the observed value than was initially given by the constant probability distribution will be seen as significantly confirmed.

\par Steven \citet{Weinberg1987} resorted to a similar kind of confirmation when he appealed to anthropic constraints (i.e., constraints related to the possibility for life to exist) in order to explain the value of the cosmological constant (i.e., the vacuum energy density $\rho_V$), left indeterminate by existing theories. Anthropic considerations provide bounds on $\rho_V$, which can be neither too large (because then galaxies could not form) nor too small (because then observers would not have time to arise before the universe recollapses). However, the anthropically allowed range is quite large, and, in the absence of further theories, it leaves us with many equipossible values for $\rho_V$. Weinberg further argued that anthropic considerations may have a stronger, predictive role. The idea is that we should conditionalize the probability of different values of $\rho_V$ on the number of observers they allow: the most likely value of $\rho_V$ is the one that allows for the largest number of galaxies (taken as a proxy for the number of observers).\footnote{This assumption is contentious (see, e.g., \citep{Aguirre2001} for an alternative proposal).} The probability measure for $\rho_V$ is then as follows:
\[
\ud p(\rho_V)=\nu(\rho_V)\cdot p_\star(\rho_V)\ud\rho_V,
\]

\noindent where $p_\star(\rho)\ud\rho_V$ is the prior probability distribution, and $\nu(\rho_V)$ the average number of galaxies which form for $\rho_V$. By assuming that there is no known reason why the likelihood of $\rho_V$ should be special at the observed value, and because the allowed range of $\rho_V$ is very far from what we would expect from available theories, Weinberg argued that it is reasonable to assume that the prior probability distribution is constant within the anthropically allowed range, so that $\ud p(\rho_V)$ can be calculated as proportional to $\nu(\rho_V)\ud\rho_V$ \citep[][2]{Weinberg2000}. Moreover, by assuming that we are typical observers (and thereby adopting what Alexander \citet{Vilenkin1995} called a `principle of mediocrity'), Weinberg predicted that the observed value of this parameter should be close to the mean of the anthropically allowed values. The initial uniform distribution is turned into a prediction, a sharply peaked distribution around a preferred value. Later observations of the cosmological constant have been taken to vindicate this reasoning \citep[see, e.g.,][]{Weinberg2007}.\footnote{The median value of the distribution obtained by such anthropic prediction is about 20 times the observed value $\rho_V^{\text{obs}}$, whereas predictions based on existing theories are 120 orders of magnitude higher than the observed value \citep{Pogosian2004}.}

John \citet{Norton2010} criticized the probabilistic representation of ignorance on which such arguments rest. He objects to the claims that 1) there can be a probability distribution over admissible values of some physical parameters when in fact those are left \emph{indeterminate} by existing theories, and more strenuously that 2) the observed value has \emph{low probability} instead of a \emph{neutral} probability. Norton has suggested that the inability to distinguish between neutral and disconfirming  evidence is the sign a fundamental flaw of Bayesianism, originating from the requirement of additivity. Therefore, for him, only a radically different inductive logic can adequately represent ignorance.

\par Norton's challenge is valid against the Bayesian who knows too much (i.e., who represents ignorance or indifference with a single uniform probability distribution). But there are other Bayesians.

\section{A Bayesian Failure?}
\label{sec:1}

\subsection{Neutral vs. Disconfirming}\label{Bayes}
\label{sec:2}

Before I discuss in more detail what a representation of neutral degree of belief---i.e., a representation of indifference or ignorance---could look like, let me rephrase the problem under consideration in Bayesian terms.

The Bayesian approach to epistemology can be characterized as follows \citep[see, e.g.,][281-282]{Joyce2010}:
\begin{list}{-}{}
\item Belief is not all or nothing. One can assign degrees of belief to propositions.
\item These degrees obey the laws of probabilities.\footnote{
(1) For a probability function $p$, $\forall \alpha, p(\alpha)\geq0$; (2) if $\alpha$ is logically true, then $p(\alpha)=1$; (3) additivity: if $\alpha,\beta$ are incompatible ($p(\alpha\&\beta)=0$), then $p(\alpha\lor\beta)=p(\alpha)+p(\beta).$ It follows from these laws that $\forall\alpha, p(\alpha)+p(\neg \alpha)=1$.
\label{laws}
}
\item Learning implies updating an initial degree of belief (called prior) to obtain a posterior. Updating a prior for hypothesis $H$ after acquiring evidence $E$ will involve taking conditional probabilities and applying Bayes's theorem:
\[
p(H|E)=\frac{p(H)\cdot p(E|H)}{p(E)},
\]
where $p(E|H)$ denotes the probability of $E$ conditional on $H$ (i.e., the probability of $E$ given $H$).
\item Rational agents use their graded beliefs to choose actions with higher expected value.
\end{list}

A controversial but usual element of probability theory, the principle of indifference claims that one must assign the same probability value to equipossible events, or events about which we are equally ignorant or indifferent. If there are many such events, additivity dictates that each of them be assigned a uniform, low probability value, which is equivalent to saying that each event is improbable (or, equivalently, that their negation is probable). 

We can now see how to rephrase the problematic arguments considered in \S1:
\begin{enumerate}
\item The value of a parameter $k$ is left indeterminate by our background knowledge. According to the principle of indifference, that indeterminateness is represented by a constant probability distribution widely spread over the admissible values of $k$, each of which having low probability $p(k|B)$.
\item A theory $T$ makes the observed value $k_{\text{obs}}$ much more probable:
\[
p(k_{\text{obs}}|T\&B)>>p(k_{\text{obs}}|B).
\]
\item According to Bayes's theorem, we then have
\[
\dfrac{p(T|k_{\text{obs}}\&B)}{p(T|B)}=\dfrac{p(k_{\text{obs}}|T\&B)}{p(k_{\text{obs}}|B)}>>1.
\]
In other words, observing $k_{\text{obs}}$ lends strong support to $T$.
\end{enumerate}

For Norton, this confirmation is unwarranted. It is based on a flaw of Bayesianism itself---namely, it has no ability to represent neutrality due to additivity \citep[501-502]{Norton2010}. Assigning a definite low probability value to a proposition about which we are ignorant is then \emph{turning ignorance into improbability}. It comes down to conflating disconfirming evidence ($p(H|E)<<1$) with neutral evidential support ($p(H|E)=p(\neg H|E)$).

\subsection{A Non-Bayesian Notion of Neutral Support?}\label{Norton_model}
\label{sec:3}

\citet{Norton2007b,Norton2008a,Norton2010} introduced the following criteria for a candidate for representation of neutral evidential support (or indifference, or ignorance):
\begin{list}{-}{}
\item it cannot be additive (and therefore does not obey the laws of probability),
\item it cannot be represented by the degrees of a one-dimensional continuum, such as the reals in $[0,1]$,
\item it must be invariant under redescription,\footnote{The invariance under redescription only requires that the probability value that corresponds to neutral support for a same event must not depend on how this event is described. For instance, in the example given above in \S 1, book length was given in terms of number of words and could be redescribed in terms of number of pages or lines.}
\item it must be invariant under negation: if we are ignorant or indifferent as to whether or not $\alpha$, we must be equally ignorant as to whether or not $\neg\alpha$.
\end{list}

It is clear that usual Bayesianism cannot meet all these criteria. We should then look for another framework for an inductive logic that would allow one not only to express ignorance and indifference, but also to compare credences and carry out inferences and confirmation. Norton confesses that he ``know[s] of no adequate theoretical representation'' of such a framework \citep[504]{Norton2010}. He then simply refers to this representation of neutral support as `$I$', for `indifference' or `ignorance', with the following properties expressed in terms of a (non-probabilistic) credal function $p$:
\begin{list}{-}{}
\item $\forall \alpha, p(\alpha)=I\rightarrow p(\neg\alpha)=I$ (invariance under negation),
\item $\forall \alpha_1,\alpha_2 \text{ mutually exclusive (but $\alpha_1\lor\alpha_2\not\equiv\top$)}, p(\alpha_1)=p(\alpha_2)=I\rightarrow p(\alpha_1\lor\alpha_2)=I$ (non-additivity).\footnote{Strictly speaking, it is not entirely appropriate to define this condition in terms of additivity. For a representation of credence to be `non-additive' in the sense of interest to Norton here, it has to fulfill the following condition: $\forall \alpha,\beta$ incompatible propositions about which we are completely indifferent or ignorant, we can have neither $p(\alpha\lor\beta)>p(\alpha)$ nor $p(\alpha\lor\beta)>p(\beta)$.\label{non-additive}}
\end{list}
This framework must also preserve the values $p(\top)=1=1-p(\bot)$.\footnote{$\top$ is an unconditionally true statement, and $\bot$ an unconditionally false one.}

\par This set of criteria for a representation of neutral evidential support is compelling. Because of this conflict with additivity, representing neutrality is a serious challenge for Bayesianism. But I will show that we need not abandon Bayesianism altogether, and that enriched versions of it already in use can satisfy these criteria to a certain extent.

\section{A Bayesian Notion of Neutral Support}
\label{sec:4}

\subsection{Bayesian Credences Need Not Have Sharp Values}
\label{sec:5}

It has been argued \citep[see, e.g.,][]{Levi1974,Walley1991,Joyce2010} that Bayesian credences need not have sharp values, and that there can be imprecise probabilities.\footnote{`Imprecise credence' is more appropriate than `imprecise probability' since it does not necessarily obey the laws of probability. Here I nevertheless use both expressions interchangeably, as is done in the literature.} The difficulty to assign sharp values to credences was already raised by \cite{Kyburg1978}, who saw this as psychologically unrealistic. An imprecise probabilities model recognizes ``that our beliefs should not be any more definitive or unambiguous than the evidence we have for them.'' \citep[320]{Joyce2010} 

\par It is possible to reject a precise Bayesian model in favor of a less exact one, in which credences are not well-defined but allow for \emph{imprecise values}. Joyce defended an imprecise model in which credences are not represented merely by a range of values, but rather by a \emph{family} of (probabilistic) credence functions. In this imprecise probability model,

\begin{quote}
\begin{enumerate}
\item a believer's overall credal state can be represented by a family $C$ of credence functions [$c_i$] (\ldots). Facts about the person's opinions correspond to properties common to all the credence functions in her credal state.
\item If the believer is rational, then every credence function in $C$ is a probability.
\item If a person in credal state $C$ learns that some event $D$ obtains (\ldots), then her post-learning state will be $C_D=\left\{c(X|D)=c(X)\dfrac{c(D|X)}{c(D)}: c\in C\right\}.$
\item A rational decision-maker with credal state $C$ is obliged to prefer one action $\alpha$ to another $\alpha^*$ when $\alpha$'s expected utility exceeds that of $\alpha^*$ relative to every credence function in $C$.\label{4} \citep[288]{Joyce2010}
\end{enumerate}
\end{quote}

\noindent In other words, in this imprecise probability model, a credal state can be represented by a family of functions that behave as usual Bayesian, probabilistic credence functions. Joyce also offers an analogy that illustrate this model: the overall credal state $C$ acts as a committee whose members (each being analogous to a credence function $c_i$) are rational agents who do not all agree with each other and who all update their credence in the same way, by conditionalizing on evidence they all agree upon. With this analogy, the properties of the overall credal state $C$ correspond to those common to all the committee members.

There are several criteria for decision-making with imprecise probabilities between two propositions. Depending on the criterion chosen, one will prefer an event to another event if
\begin{list}{-}{}
\item it has maximum lower expected value ($\Gamma-$minimax criterion),
\item it has maximum higher expected value ($\Gamma-$maximax),
\item it has maximum expected value for all distributions in the credal set (maximality),
\item it has a higher expected value for at least one distribution in the credal set ($E-$admissibility), or 
\item its lower expected value on all distributions in the credal set is greater than the other event's highest expected value on all distributions (interval dominance).\footnote{This list is not exhaustive. It is beyond the scope of this paper to compare and assess these criteria. See \citep[\S8]{Troffaes2007,Augustin2014} for reviews.}
\end{list}

\par This model allows one to simultaneously represent sharp and imprecise credences, but also comparative probabilities. It can accommodate sharp credences, for which the usual condition of additivity holds. But it can also accommodate less sharply defined relationships when credences are fuzzy. It does so by means of a family of credence functions, each of which is a Bayesian function that obeys the laws of probability.

\subsection{Neutral Support with Imprecise Credences}
\label{sec:6}

\par This imprecise model allows one to distinguish two notions of neutral support: distinguish stochastic independence (true independence of evidence for a given hypothesis) and unknown interaction (unknown dependence).

With a single (precise) probability distribution, two variables $\alpha_m$ and $\alpha_n$ in appropriate algebras are stochastically independence if $p(\alpha_m |\alpha_n )=p(\alpha_m)$ whenever $p(\alpha_n )>0$. Different concepts have been suggested to extend the notion of stochastic independence of two variables $\alpha_m$ and $\alpha_n$ to the imprecise model \citep[see][]{Cozman2012}, among which:
\begin{list}{-}{}
\item complete independence, if stochastic independence of $\alpha_m$ and $\alpha_n$ obtains for each distribution $c_i\in C$,\footnote{As discussed in \citep{Cozman2012}, this definition violates convexity.}
\item confirmational irrelevance, if $C(\alpha_m|\alpha_n)=C(\alpha_m)$,
\item epistemic irrelevance, if stochastic independence obtains for the lowest expectation among all functions $c_i\in C$, or the related, symmetric concept of epistemic independence (if $\alpha_m$ is epistemically irrelevant to $\alpha_n$ and $\alpha_n$ to $\alpha_m$).
\end{list}

In contrast, unknown interaction between two variables $\alpha_m$ and $\alpha_n$ can be represented by a credal set that contains credence functions that differ on the correlation between $\alpha_m$ and $\alpha_n$. Such a credal set would include credence functions $c_i$ that are such that $c_i(\alpha_m \& \alpha_n)\leq c_i(\alpha_m) \cdot c_i(\alpha_n)$ and others functions $c_j$ such that $c_j(\alpha_m \& \alpha_n)>c_j(\alpha_m) \cdot c_j(\alpha_n)$. Following the committee analogy once again, in a state of unknown interaction between $\alpha_m$ and $\alpha_n$, jury members disagree with each other about that interaction. The overall credal state of the committee neither favors nor dismisses any of the opinions of its jury members; it can only express a lack of agreement.

\par The case of complete ignorance or indifference is of particular interest for the anthropic argument presented in \S1: we want to represent \emph{complete ignorance or indifference} as to what value a certain physical parameter should have. With the imprecise model, recalling the committee analogy, complete ignorance as to whether or not $\alpha$ (i.e., complete indifference between which of $\alpha$ or $\neg\alpha$ is more likely) can be represented by a committee in which there is \emph{no agreement} among its members about whether or not $\alpha$ is more or less probable than $\neg\alpha$. This notion of neutral support is analog to Norton's requirement that, in order to express neutral support for an event $\alpha$, a measure of belief about $\alpha$ be equal to a corresponding measure of \emph{disbelief} about $\alpha$ \citep{Norton2007b}. In the context of credal sets, this requirement is fulfilled, for example, by a committee composed as follows:
\begin{quote}
for every jury member and for every contingent event $\alpha$, there exists a jury member who has as much disbelief in $\alpha$ as another has belief in it.
\end{quote}
Or, to put it in terms of credence functions,
\begin{quote}
if $C$ is a credal state representing a family of credence functions $c$ about contingent events $\alpha$, if $\forall c,\alpha$, $\exists\, c'\in C$ such that $c'(\alpha)=1-c(\alpha)$, then $C$ corresponds to a state of ignorance.\footnote{It may be surprising to read that, for a certain proposition, a credal set that gives the set of values $\{0.49,0.51\}$ is a better representation of ignorance than one that gives the set of values $\{0.1,0.8\}$. It is here a representation of ignorance that gives as much support to a proposition as it does to its negation that this criterion captures.}
\end{quote}\label{oneminus}

If all the jury members are rational (i.e., if all the credence functions are probability functions), we must have, by definition, $C(\top)=1=1-C(\bot).$\footnote{i.e., $\forall c\in C, c(\top)=1=1-c(\bot).$}

\par Such a credal state about a given event $\alpha$ can then be updated by conditionalizing upon new evidence, taken as such by all the jury members (i.e., all the credence functions $c\in C$). Thereby, after new evidence $E$ is gathered, the range of values taken by all the credence functions of $C$, $c(\alpha|E)$, is susceptible to change, and so is the overall credal state regarding $\alpha$.

A trivial but extreme example of representation of \emph{complete} ignorance by means of a set of credal functions is the set $\textgoth{I}$ of \emph{all} possible probability distributions. For any proposition $\alpha$ about which are completely ignorant, that representation of ignorance would give us $C(\alpha)=[0,1]$. In case of complete ignorance, excluding possible probability distributions compatible with our evidence is ``pretending to have information [we do] not possess.'' \citep[170]{Joyce2005} This proposal meets all of Norton's criteria for a representation of ignorance (see \emph{infra}, \S~\ref{Norton_model}).

There is however a good reason not to be content with such an extreme representation of ignorance. Indeed, in that set \textgoth{I} of all possible probability distributions will be extremely sharp probability distributions that require an unreasonably large---or even infinite---number of updatings before they can yield posteriors distributions that are significantly different \citep{Rinard2013}. Such distributions in \textgoth{I} are said to be dogmatic, and consequently the whole set \textgoth{I} is dogmatic. A representation of complete ignorance \textgoth{I}, and generally any vacuous prior, entails a vacuous posterior. This should prevent such a set from being used in an inferential process in which we may hope to move \emph{away} from a state of ignorance after a certain number of iterations of Bayesian updating. This representation of ignorance by means of a family of credal functions, although it satisfies Norton's criteria for ignorance, is \emph{incompatible with learning}. That is why imprecise statisticians, who are interested in inferential processes, prefer to deal with `near-ignorance' (i.e., broad credal intervals smaller than [0,1]) rather than complete ignorance, thereby ruling out dogmatic priors \citep[see][\S7.3.7]{Moral2012,Augustin2014,Walley1991}.

By way of example, let us represent our near-ignorance about three mutually exclusive propositions $\alpha_1, \alpha_2, \alpha_3$ with a credal set $C=\{c_i\}$ consisting of probability functions defined as follows:

\begin{center}
    \begin{tabular}{l|*{3}{c}r}
     & $\alpha_1$ & $\alpha_2$ & $\alpha_3$ \\ \hline
    $c_1(\alpha_i)$ & 0.9 & 0.05 & 0.05 \\
    $c_2(\alpha_i)$ & 0.05 & 0.9 & 0.05 \\
    $c_3(\alpha_i)$ & 0.05 & 0.05 & 0.9 \\
\end{tabular}
\end{center}

\noindent and for all $i,j\in\{1,2,3\}$, for all $x\neq \alpha_i$ or their disjunctions, $c_j(x)=0$.

In this credal set, no value of $\alpha_m$ is preferred to the others (i.e., no value of $\alpha_m$ is favored by all three distributions $c_i\in C$); and for all $\alpha_m, \alpha_n$, $m\neq n$, $C(\alpha_m)=C(\alpha_n)$. A constant, unique probability distribution could express this as well. But this credal set tells us more than that, namely that, for all $m$, no proposition $\alpha_m$ is preferred to its negation (i.e., no value of $\alpha_m$ is favored over its negation by all three distributions $c_i \in C$).

\par This example does not satisfy the characterization of ignorance proposed earlier. However, we can see that, with a few amendments, this set suffices to represent indifference about any of these propositions $\alpha_i$, and can meet the requirements for a representation of neutral support (see above \S\ref{Norton_model}):

\begin{list}{-}{}
\item it is not a sharp value in $[0,1]$,
\item it can be defined so as to be invariant under redescription: $\forall \alpha, a$ ($a$, redescription of $\alpha$), if $C(\alpha)$ represents a state of ignorance, so will $C(a)$,\footnote{If the functions in this set are described as Dirichlet distributions, then this criterion will be satisfied \citep[see, e.g.,][]{DeCooman2009}.\label{Dir1}}
\item it is invariant under negation (no proposition $\alpha$ is preferred to its negation).
\item we do have $C(\alpha_1\lor\alpha_2\lor\alpha_3)=C(\top)=1=1-C(\bot)$.
\end{list}

\noindent The criterion of non-additivity (see \emph{infra}, note~\ref{non-additive}) cannot be satisfied in a trivial manner. In the example above, we have $\forall i,m,n\in\{1,2,3\}, m\neq n, c_i(\alpha_m\lor\alpha_n)>c_i(\alpha_m)$. That is also true of the bounds of the credal intervals (they will not be non-additive). But we can prevent that by adding to our set credal functions $c_k$ such that $c_k(\alpha_1\lor\alpha_2)=c_k(\alpha_1)$ and so on. That is, we can add to our initial set three other functions $c_4,\,c_5,$ and $c_6$ so as to have $\forall i\in\{1,2,3\}, \exists\, c_j\in C, j\in\{4,5,6\}$ such that $c_j(\alpha_i)=0$. But the newly added functions are not reasonable, since if we agreed that the propositions $\alpha_i$ are contingent, then no function ruling them out completely should be accepted in our credal set. 

However, adopting certain conventions can mitigate the effects of additivity. If among the criteria for decision-making with credal sets introduced earlier in \S~\ref{sec:5} we choose that of interval dominance, then our representation of credence is non-additive in this example (i.e., for all $m,n\in\{1,2,3\}$, $\alpha_m\lor\alpha_n$ is not preferred to $\alpha_m$). However, one might argue that interval dominance is often not a desirable criterion. It is a demanding criterion that allows one to express a preference between two propositions only if one is unambiguously better than the other. This criterion is arguably not fined-grained enough to help us for most of the inferences we are likely to encounter. Other, often more desirable decision-making criteria, however, will result in additive imprecise credences. With other decision rules, we could under circumstances circumvent additivity by adopting threshold values, beyond (respectively below) which all values are considered to be equally confirmatory (respectively disconfirmatory). For instance, in the set given above, if we consider that any value below 0.1 is equally disconfirmatory and any value beyond 0.9 is equally confirmatory, then we lose all additivity. 

We could also adopt a different strategy that does not involve such conventions. For any proposition $\alpha$ about which we are ignorant (whether the $\alpha_i$ or their Boolean combinations), we can define a credal set $C$ such that $C(\alpha)=C(\neg \alpha)$. The imprecise model allows us to treat equipossible propositions in the same way, which does not mean that they must receive a same probability value or be represented by the same credal set. By `sameness of treatment' of contingent, equipossible events about which we are ignorant or indifferent, I mean that our credal state of ignorance about them would be modeled in the same manner, by a credal set having the same desired properties. We will come back to this in the next section.

With the imprecise model, not one single set of functions or one set of rules to define it will be suited to truly represent ignorance in all situations, unless we are ready to represent a state of complete ignorance by the undesirable and unreasonable set \textgoth{I} of all possible credal functions. But this model allows one to define sets that do not favor any of the propositions about which we are ignorant and that is suited to a particular ensemble of propositions under consideration. It does so by means of Bayesian credence functions, which allows for our credal state to be updated and evolve. All that this requires is that we do not demand that agent's credences have sharp values.

\subsection{Norton's Objections}
\label{sec:7}

\subsubsection{Interpretative objections}
\label{sec:8}

\citet{Norton2007b,Norton2007a,Norton2008a} has formulated several objections to the imprecise model. He has expressed a general discomfort with what he considers to be an inadequate approach, ``an attempt to simulate an inherently nonadditive logic with an additive measure, rather than to seek the logic directly.'' \citep[504, note 4]{Norton2010}. But as indirect or contrived as this method might seem, it is successfully applied in statistical analysis \citep[see, e.g.,][]{Walley1991,Augustin2014}. Yet Norton has raised more pointed criticisms of the imprecise model, about the very question of representing ignorance:

\begin{quote}
the representation [of ignorance by sets of probability functions] is not literally correct. That is, ignorance is not the maintaining of all possible beliefs at once; it is the maintaining of none of them. So we should regard the device of convex sets as a way of simulating ignorance through a convenient fiction.\footnote{This remark also applies to non-convex sets.} \citep[\S4.2]{Norton2007a}
\end{quote}

Or elsewhere:

\begin{quote}
The sort of ignorance I seek to characterize is first order ignorance; it is just not knowing which is the true outcome; not a second order uncertainty about an uncertainty. \citep[\S6.2]{Norton2007b}
\end{quote}

Such criticism is in fact not specific to the question of representing ignorance. It is aimed more generally at the use of \emph{several} credal functions in order to represent a \emph{unique} credal state.\footnote{The following passage makes it clear that Norton thinks of the use of a set of probability functions as allowing the simultaneous representation of several states of belief: ``the use of sets renders ignorance as a second order sort of belief. We allow that many different belief-disbelief states are possible. We represent ignorance by presenting them all, in effect saying that we don’t know which is the pertinent one.'' \citep[\S6.2, 248]{Norton2007b}} But it does not apply to the imprecise model we have considered here, and, in general, proponents of imprecise probabilities need not endorse this view. Indeed, this model does not allow ``the maintaining of all possible beliefs at once.'' Even though a credal set may be comprised of several credal functions, the agent's credence it represents is unique; its properties are those that are common to all the credal functions in that set. If all possible beliefs could be held at once---or rather, if no particular belief can be preferred to any other---an agent's credence would not be multiple, it would just \emph{not be any} of these particular beliefs.

\subsubsection{Indifference and self-duality}
\label{sec:9}

\par Another criticism more specifically aimed at the issue of representing ignorance or indifference deserves a closer examination. \citet{Norton2007b} argues that a representation of ignorance should satisfy what he calls a condition of self-duality. The dual of a measure of belief is a measure of disbelief. If we are ignorant about a proposition $\alpha$, our degree of belief that $\alpha$ should not be different from our degree of disbelief that $\alpha$: ``the state [of complete ignorance] is self-dual in its contingent propositions (\ldots).'' \citep[\S 6.1, 247]{Norton2007b} This follows from the requirement of invariance under negation for a representation of neutral support.

We saw earlier in \S~\ref{sec:6} that no single probability function (or in general no measure, additive by definition) can meet this criterion.\footnote{Unless we are dealing with only two mutually exclusive propositions.} Any probability function will necessarily express a (non-strictly) increasing belief as we go from propositions to their logical consequences (e.g., for any functions $c$ in a credal set, $\alpha_1\vdash\alpha_1\lor\alpha_2\rightarrow c(\alpha_1\lor\alpha_2)\geq c(\alpha_1)$). On the other hand, a measure of disbelief should be (non-strictly) \emph{decreasing} as we go from propositions to their logical consequences. Consequently, no probability function can simultaneously express our belief and our disbelief in a certain proposition, even if we consider only contingent propositions. This, however, is not necessarily true of credal sets. We also saw in \S~\ref{sec:6} that if decision-making is based on interval dominance, then imprecise credences need not be additive for contingent propositions. Consequently, an imprecise credence represented by a set of functions as the one given in \S~\ref{sec:6} can simultaneously represent belief and disbelief (at least for contingent propositions).

On Norton's account \citep[][\S~3.2,6.2]{Norton2007b}, a credal set $C$ representing ignorance is self-dual if it contains probability functions $c_i$ and their duals $c'_i$ such that $\forall c_i\in C, c'_i$ is such that $\forall \alpha, c_i(\alpha)=c'_i(\neg\alpha)$ and $c_i(\neg\alpha)=c'_i(\alpha)$. It is clear that we cannot have such a set. Indeed, if we have
\begin{center}
$\forall c\in C, \exists\, c'\in C \text{ such that } \forall \alpha, c'(\alpha)=1-c(\alpha)$,
\end{center}
 then if $\alpha\equiv\top$ and $c$ is a probability function, we would get
\begin{center}
$c'(\alpha\equiv\top)=1-c(\top)=0,$
\end{center}
and then $c'$ would break the second law of probability (see infra note \ref{laws}). Norton restricts the requirement of self-duality to contingent events only, which are the only relevant ones when we want to define a representation of ignorance or indifference. But even if we forget about logical truths and contradictions, additivity will forbid such a self-dual set. Indeed, for two propositions $\alpha$ and $\beta$ mutually exclusive, we do not get $\forall i, c_i'(\alpha\lor\beta)=1-c_i(\alpha\lor\beta)=c_i'(\neg(\alpha\lor\beta)).$ If it were the case, we would obtain
\begin{align*}
c_i'(\alpha\lor\beta)&=c_i(\neg(\alpha\lor\beta))\\
		&=c_i'(\alpha)+c_i'(\beta)\\
		&=c_i(\neg\alpha)+c_i(\neg\beta)
\end{align*}

For $i$ such that $c_i(\alpha)<<1$, $c_i(\neg\alpha)\simeq 1$ and we may have $c'_i(\alpha\lor\beta)=c_i(\neg\alpha)+c_i(\neg\beta)>1$, which would prevent $c_i'$ from being a probability function. If this set is not self-dual for $\alpha$ and $\beta$ mutually exclusive, it cannot be self-dual for all $\alpha$ and the credal state as I have defined it cannot be self-dual.

In contrast, as we saw above in \S~\ref{oneminus}, for any set of contingent propositions, the imprecise model allows us to define sets that represent our indifference or ignorance. The proposal for a representation of neutral support we saw above in \S~\ref{oneminus}, according to which if $C$ is a credal state representing a family of credence functions $c$ about contingent events $\alpha$, $C$ corresponds to a state of indifference if $\forall c,\alpha$, $\exists\, c'\in C$ such that $c'(\alpha)=1-c(\alpha)$, differs from Norton's criterion of self-duality (note the different placement of the universal quantifier on $\alpha$). The imprecise model as I have considered it here does not guarantee us than once a credal set is defined so as to represent indifference about a set of propositions, it will necessarily adequately represent our indifference about all their Boolean combinations, nor, in general, that it will adequately provide neutral support to any other proposition we are indifferent about. This model does not suggest either that the same rules used to define a credal set representing indifference should apply to all situations (this will be illustrated below in \S~\ref{sec:10}).

If we consider only contingent propositions, and if decision-making is based on interval dominance, then with imprecise probabilities we have at our disposal a representation of ignorance or indifference that shares with a self-dual representation the relevant properties for a representation of neutral support, namely the ability to simultaneously represent belief and disbelief. With other decision rules (e.g., $\Gamma$-minimax), a self-dual representation of ignorance or indifference could only express nothing less than \emph{complete} ignorance. We saw in \S~\ref{oneminus} that such a credal set, \textgoth{I}, exists, but we also saw that it is incompatible with learning. The less demanding representations of ignorance or indifference I have considered here have over a self-dual representation a clear expressive advantage. We saw that the imprecise model allows one to distinguish between stochastic independence, epistemic irrelevance, unknown interaction, or ignorance or indifference about the value of a parameter. It can fulfill criteria of a representation of ignorance better than what a single probability function can do, yet it does so by means of probability functions, each of which can subsequently be modified following Bayes's rule. Consider for instance the 3-function example from \S~\ref{oneminus}. Now assume that the probability of $\alpha_i$ in that example corresponds to the probability of drawing the numbered ball $B_i$ from an urn (with replacement between each draw). Assume further that we are interested in determining a bias---of our urn or the balls---in this drawing. How should we represent our initial state of ignorance in a way that does not prevent us from eventually finding a value after a sufficient iteration of Bayesian updating? It is unclear how Norton's non-probabilistic, self-dual representation of complete ignorance, `$I$', can evolve, whereas it is possible for our 3-function credal set to yield such a result eventually.\footnote{See \citep{Piatti2009,Moral2012} for a discussion about the conditions on a near-ignorance credal set required for learning. These requirements favor the use of Dirichlet distributions.\label{Dir2}} Notwithstanding the lack of self-duality, the ability for credal sets representing indifference or ignorance to be updated is a desirable feature that makes an imprecise representation of indifference a more interesting element of inductive logic than Norton's self-dual measure $I$.

The kind of representation of ignorance that Norton seeks is part of a larger inductive logic yet to be carried out. It is in the light of the search for a unique representation of our credences that Norton imposes this criterion of self-duality of complete ignorance, applicable to \emph{any} proposition about which we are ignorant. In this context, it is a plausible requirement. But it is a very strong condition that can leave one wondering how a unique representation as the one Norton proposes---which would yield the same value `$I$' whatever the event about which we are ignorant---can be used in a fruitful inductive process. Norton offers no compelling reason why his demanding notion of self-duality should be required of alternative representations of ignorance or indifference such as the ones the imprecise model affords us.

\section{Dissolving Cosmic Confusions}
\label{sec:10}

In order to emphasize how ill-equipped Bayesianism (or any other inductive framework based on additive measures) is in order to deal with ignorance, and to show its inability to avoid unwarranted conclusions, \citet{Norton2010} invited us to consider two instances of `cosmic confusion.' These are anthropic reasoning (see infra, \S~\ref{intro}) and the Doomsday argument. In both cases, we draw conclusions from a \emph{lack of knowledge}. We can see how the use of imprecise credences can dissolve what Norton referred to as `cosmic confusions' in a way no unique probability distribution can.\footnote{I have given a more detailed presentation of the problems and solutions discussed in this section in \citep{Benetreau-Dupin2015}.}

The Doomsday argument is a family of arguments about humanity's likely survival (see, e.g., \citep[][\S6-7]{Bostrom2002}, \citep{Richmond2006} for reviews). It allows one to compare the likelihood of two scenarios about humanity's survival or even make a prediction about the end date for humanity based only on the assumption that our place on humanity's timeline is random. Some versions of this argument have been addressed within the framework of orthodox Bayesianism so as to block any conclusion we could draw from these assumptions alone. But a variant of this argument, based on the assumption that we have a random birth rank among all humans \citep{Gott1994}, cannot be dissolved without appealing to imprecise credences. The argument goes as follows. Let $r$ be my birth rank (i.e., I am the $r^{\text{th}}$ human to be born), and $N$ the total number of humans that will ever be born.

\begin{enumerate}
\item Assume that there is nothing special about my rank $r$. Following the principle of indifference, whatever $r$, the probability of $r$ conditional on $N$ is $p(r|N)=\dfrac{1}{N}$.
\item Assume the following improper prior\footnote{As \citet{Gott1994} recalls, this choice of prior is fairly standard (albeit contentious) in statistical analysis. It is the Jeffreys prior for the unbounded parameter $N$, such that $p(N)\ud N\propto\ud \ln N\propto\dfrac{\ud N}{N}$. This means that the probability for $N$ to be in any logarithmic interval is the same. This prior is called improper because it is not normalizable, and it is usually argued that it is justified when it yields a normalizable posterior.}\label{Gott_Jeffreys} for $N$: $p(N)=\dfrac{k}{N}$. $k$ is a normalizing constant whose value does not matter.
\item This choice of distributions $p(r|N)$ and $p(N)$ gives us the prior distribution $p(r)$:
\[
p(r)=\int_{N=r}^{N=\infty}p(r|N)p(N)\ud N=\int_{N=r}^{N=\infty}\dfrac{k}{N^2}\ud N=\dfrac{k}{r}.
\]
\item Then, Bayes's theorem gives us
\[
p(N|r)=\dfrac{p(r|N)\cdot p(N)}{p(r)}=\dfrac{r}{N^2},
\]
which favors small $N$ and allows us to make an estimate for $N$ at any confidence-level.
\end{enumerate}

This result should strike us as surprising: we should not be able to learn something from nothing! Indeed, according to that argument, we can make a prediction for $N$ based only on knowing our rank $r$ and on \emph{not} knowing anything about the probability of $r$ conditional on $N$, i.e., on being indifferent---or equally uncommitted---about any value it may take.

In this argument, any choice of prior probability distribution will result in a prediction for $N$, at any confidence-level. However, if our prior ignorance or indifference about $N$, $C(N),$ is represented by a credal set containing an \emph{infinity} of credal functions, $\{c:c \in C\}$, each normalizable, defined on $\mathbb{N}^{>0}$, and such that $\forall c \in C,\lim_{N\rightarrow\infty}(c (N))=0$ (e.g., a family of Pareto distributions), then the resulting prediction for $N$ diverges.  In other words, this imprecise representation of prior credence in $N$, reflecting our ignorance about $N$, does not yield any prediction about $N$. Without the possibility for my prior credence to be represented not by a single probability distribution, but instead by an \emph{infinite set} of probability distributions, I cannot avoid obtaining an arbitrarily precise prediction.

In the case of the cosmological constant problem (see infra \S~\ref{intro}), representing our prior ignorance or indifference about the value of the vacuum energy density $\rho_V$ by an imprecise credal set can limit, if not entirely dissolve, the appeal of anthropic considerations. As we saw earlier, \citet{Weinberg1987} argued that, in the absence of useful theoretical background, it was reasonable to assume a constant, uniform prior probability distribution for $\rho_V$ within the anthropically allowed range, and then conditionalize on the number of observers each value of $\rho_V$ would allow for. With the imprecise model, a state of indifference between different values of $\rho_V$ within the anthropically allowed range can be expressed by a set of probability distributions $\{c_\star:c_\star\in C_\star\}$, all of which normalizable over the anthropic range and such that $\forall\rho_V, \exists\, c_{\star i}, c_{\star j}\in C_\star$ such that $\rho_V$ is favored by $c_{\star i}$ and not by $c_{\star j}$.\footnote{For reasons expressed earlier in footnotes \ref{Dir1} and \ref{Dir2}, this should preferably be done by means of Dirichlet distributions.} It is in principle possible to define this prior credal set so that for any value of $\rho_V$, the lowest expectation value among the the posteriors is lower than the highest expectation value among the priors. If then we adopt interval dominance as a criterion for decision-making, then no observation of $\rho_V$ will be able to lend support to our anthropic prediction.

As we saw earlier, one may object to the adoption of interval dominance in such a case. This choice of demanding decision rule could be motivated by the fact that we have no plausible alternative theoretical framework to the anthropic argument. In this context, it can be reasonable to agree to increase one's credence about the anthropic explanation \emph{only if} it does better than any other yet unknown alternative might have done. Nonetheless, if we adopt other decision rules, it is possible with the imprecise model to construct prior credal sets that define a large interval over the anthropic range such that the confirmatory boost obtained after observing $\rho_V$ is not nearly as vindicative as it is with a single, uniform distribution.

\section{Conclusion}
\label{sec:11}

\citet{Norton2010} has correctly argued that representing neutrality with a broadly spread single probability distribution amounts to conflating ignorance with improbability. He has shown how this leads to unwarranted confirmations. I here claim that a credal state of ignorance should best be represented by an imprecise credence. With this approach, merely acquiring information about the value of a certain parameter cannot suffice to justify a sense of surprise. The imprecise model offers us a more adequate representation of neutrality and prevents prior credences from doing too much inductive work, as is illustrated by its ability to block the consequence of the Doomsday argument better than what orthodox Bayesianism can do.

\par We saw that, if we adopt interval dominance as a criterion for decision-making, it is possible for an imprecise representation of indifference to meet the criteria for a representation of neutral support put forth in \citep{Norton2007b,Norton2008a,Norton2010}. But if we adopt less demanding decision-making rules, we can still construct a representation of indifference by means of credal sets that meet these criteria to a large extent, and we can do so without compromising Bayesianism altogether. It only requires that we do not demand that credences be sharp nor that a unique representation be applicable to all cases of ignorance or indifference (i.e., that self-duality be abandoned).

\par One can see Norton's argument as emphasizing the perils of excessive structure imported from probability theory into inductive logic, and then arguing that we need to eliminate much of that structure. There are several ways to modify the mathematical structure to counter the assumption of additivity. Norton's stated motivations are not sufficient to force us to adopt the framework he advocates. The imprecise model has the advantage of allowing us to distinguish different kinds of ignorance and indifference. More importantly, it makes it possible to move out of a state of ignorance when we acquire interesting information.

\subsection*{Acknowledgements}

I am grateful to Wayne Myrvold for initial discussions, Jim Joyce and John Norton for stimulating exchanges. I am indebted to Chris Smeenk for many comments and suggestions.



%
%



\begin{thebibliography}{}

\bibitem[\protect\citeauthoryear{Aguirre}{Aguirre}{2001}]{Aguirre2001}
Aguirre, A. (2001, September).
\newblock {Cold Big-Bang Cosmology as a Counterexample to Several Anthropic
  Arguments}.
\newblock {\em Physical Review D\/}~{\em 64\/}(8), 1--12.

\bibitem[\protect\citeauthoryear{Augustin, Coolen, de~Cooman, and
  Troffaes}{Augustin et~al.}{2014}]{Augustin2014}
Augustin, T., F.~P. Coolen, G.~de~Cooman, and M.~C. Troffaes (Eds.) (2014).
\newblock {\em {Introduction to Imprecise Probabilities}}.
\newblock Wiley \& Sons.

\bibitem[\protect\citeauthoryear{Ben\'etreau-Dupin}{Ben\'etreau-Dupin}{forth.}]{Benetreau-Dupin2015}
Ben\'etreau-Dupin, Y. (\emph{Forthcoming}).
\newblock {Blurring Out Cosmic Puzzles}.
\newblock {\em Philosophy of Science\/}.

\bibitem[\protect\citeauthoryear{Bostrom}{Bostrom}{2002}]{Bostrom2002}
Bostrom, N. (2002).
\newblock {\em {Anthropic Bias: Observation Selection Effects in Science and
  Philosophy}}.
\newblock Routledge.

\bibitem[\protect\citeauthoryear{Cozman}{Cozman}{2012}]{Cozman2012}
Cozman, F.~G. (2012).
\newblock {Sets of Probability Distributions, Independence, and Convexity}.
\newblock {\em Synthese\/}~{\em 186\/}(2), 577--600.

\bibitem[\protect\citeauthoryear{de~Cooman, Miranda, and Quaeghebeur}{de~Cooman
  et~al.}{2009}]{DeCooman2009}
de~Cooman, G., E.~Miranda, and E.~Quaeghebeur (2009, February).
\newblock {Representation Insensitivity in Immediate Prediction under
  Exchangeability}.
\newblock {\em International Journal of Approximate Reasoning\/}~{\em 50\/}(2),
  204--216.

\bibitem[\protect\citeauthoryear{Gott}{Gott}{1994}]{Gott1994}
Gott, J.~R. (1994).
\newblock {Future Prospects Discussed}.
\newblock {\em Nature\/}~{\em 368\/}(March), 108.

\bibitem[\protect\citeauthoryear{Joyce}{Joyce}{2005}]{Joyce2005}
Joyce, J.~M. (2005).
\newblock {How Probabilities Reflect Evidence}.
\newblock {\em Philosophical Perspectives\/}~{\em 19}, 153--178.

\bibitem[\protect\citeauthoryear{Joyce}{Joyce}{2010}]{Joyce2010}
Joyce, J.~M. (2010).
\newblock {A Defense of Imprecise Credences in Inference and Decision Making}.
\newblock {\em Philosophical Perspectives\/}~{\em 24\/}(1), 281--323.

\bibitem[\protect\citeauthoryear{Kyburg}{Kyburg}{1978}]{Kyburg1978}
Kyburg, H. (1978).
\newblock {Subjective Probability: Criticisms, Reflections, and Problems}.
\newblock {\em Journal of Philosophical Logic\/}~{\em 7\/}(1), 157--180.

\bibitem[\protect\citeauthoryear{Levi}{Levi}{1974}]{Levi1974}
Levi, I. (1974).
\newblock {On Indeterminate Probabilities}.
\newblock {\em The Journal of Philosophy\/}~{\em 71\/}(13).

\bibitem[\protect\citeauthoryear{Moral}{Moral}{2012}]{Moral2012}
Moral, S. (2012, April).
\newblock {Imprecise Probabilities for Representing Ignorance about a
  Parameter}.
\newblock {\em International Journal of Approximate Reasoning\/}~{\em 53\/}(3),
  347--362.

\bibitem[\protect\citeauthoryear{Norton}{Norton}{2007a}]{Norton2007b}
Norton, J.~D. (2007a, October).
\newblock {Disbelief as the Dual of Belief}.
\newblock {\em International Studies in the Philosophy of Science\/}~{\em
  21\/}(3), 231--252.

\bibitem[\protect\citeauthoryear{Norton}{Norton}{2007b}]{Norton2007a}
Norton, J.~D. (2007b, May).
\newblock {Probability Disassembled}.
\newblock {\em The British Journal for the Philosophy of Science\/}~{\em
  58\/}(2), 141--171.

\bibitem[\protect\citeauthoryear{Norton}{Norton}{2008}]{Norton2008a}
Norton, J.~D. (2008).
\newblock {Ignorance and Indifference}.
\newblock {\em Philosophy of Science\/}~{\em 75\/}(1), 45--68.

\bibitem[\protect\citeauthoryear{Norton}{Norton}{2010}]{Norton2010}
Norton, J.~D. (2010).
\newblock {Cosmic Confusions: Not Supporting versus Supporting Not}.
\newblock {\em Philosophy of Science\/}~{\em 77\/}(4), 501--523.

\bibitem[\protect\citeauthoryear{Piatti, Zaffalon, Trojani, and Hutter}{Piatti
  et~al.}{2009}]{Piatti2009}
Piatti, A., M.~Zaffalon, F.~Trojani, and M.~Hutter (2009, April).
\newblock {Limits of Learning about a Categorical Latent Variable under Prior
  Near-Ignorance}.
\newblock {\em International Journal of Approximate Reasoning\/}~{\em 50\/}(4),
  597--611.

\bibitem[\protect\citeauthoryear{Pogosian, Vilenkin, and Tegmark}{Pogosian
  et~al.}{2004}]{Pogosian2004}
Pogosian, L., A.~Vilenkin, and M.~Tegmark (2004, July).
\newblock {Anthropic Predictions for Vacuum Energy and Neutrino Masses}.
\newblock {\em Journal of Cosmology and Astroparticle Physics\/}~{\em
  7\/}(005), 1--17.

\bibitem[\protect\citeauthoryear{Richmond}{Richmond}{2006}]{Richmond2006}
Richmond, A. (2006).
\newblock {The Doomsday Argument}.
\newblock {\em Philosophical Books\/}~{\em 47\/}(2), 129--142.

\bibitem[\protect\citeauthoryear{Rinard}{Rinard}{2013}]{Rinard2013}
Rinard, S. (2013).
\newblock {Against Radical Credal Imprecision}.
\newblock {\em Thought: A Journal of Philosophy\/}~{\em 2\/}(2), 157--165.

\bibitem[\protect\citeauthoryear{Troffaes}{Troffaes}{2007}]{Troffaes2007}
Troffaes, M. (2007).
\newblock {Decision Making Under Uncertainty Using Imprecise Probabilities}
\newblock {\em International Journal of Approximate Reasoning\/}~{\em 45\/}(1), 17--29.


\bibitem[\protect\citeauthoryear{Vilenkin}{Vilenkin}{1995}]{Vilenkin1995}
Vilenkin, A. (1995).
\newblock {Predictions from Quantum Cosmology}.
\newblock {\em Physical Review Letters\/}~{\em 74\/}(6), 4--7.

\bibitem[\protect\citeauthoryear{Walley}{Walley}{1991}]{Walley1991}
Walley, P. (1991).
\newblock {\em {Statistical Reasoning with Imprecise Probabilities}}.
\newblock London: Chapman and Hall.

\bibitem[\protect\citeauthoryear{Weinberg}{Weinberg}{1987}]{Weinberg1987}
Weinberg, S. (1987).
\newblock {Anthropic Bound on the Cosmological Constant}.
\newblock {\em Physical Review Letters\/}~{\em 59\/}(22), 2607--2610.

\bibitem[\protect\citeauthoryear{Weinberg}{Weinberg}{2000}]{Weinberg2000}
Weinberg, S. (2000).
\newblock {A Priori Probability Distribution of the Cosmological Constant}.
\newblock {\em arXiv preprint astro-ph/0002387\/}, 0--15.

\bibitem[\protect\citeauthoryear{Weinberg}{Weinberg}{2007}]{Weinberg2007}
Weinberg, S. (2007).
\newblock {Living in the Multiverse}.
\newblock In B.~Carr (Ed.), {\em Universe or Multiverse?}, Chapter~2, pp.\
  29--42. Cambridge University Press.

\end{thebibliography}
\end{document}